\documentclass[aps,prl,preprint,groupedaddress]{revtex4}

\usepackage{graphicx}
\usepackage{dcolumn}
\usepackage{bm}

\begin{document}


\title{Electric field Induced Patterns in Soft Visco-elastic films: From Long Waves of Viscous Liquids to Short Waves of Elastic Solids}


\author{N. Arun$^{1}$}
\author{ Ashutosh Sharma$^{2}$}
\email{ashutos@iitk.ac.in}
\author{Partho S. G. Pattader$^{2}$}
\author{Indrani Banerjee$^{2}$}
\author{Hemant M. Dixit$^{1}$}
\author{K. S. Narayan$^{1}$}
\email{narayan@jncasr.ac.in}
\affiliation {
$^{1}$Jawaharlal Nehru Centre for Advanced Scientific Research, Bangalore 560064, India.\\
$^{2}$Department of Chemical Engineering, Indian Institute of Technology, Kanpur 208016, India.}

\date{\today}

\begin{abstract}
We show that the electric field driven surface instability of visco-elastic films has two distinct regimes: (1) The visco-elastic films behaving like a liquid display long wavelengths governed by applied voltage and surface tension, independent of its elastic storage and viscous loss moduli, and (2) the films behaving like a solid require a threshold voltage for the instability whose wavelength always scales as $\sim$ 4 $\times$  film thickness, independent of its surface tension, applied voltage, loss and storage moduli. Wavelength in a narrow transition zone between these regimes depends on the storage modulus.
\end{abstract}



\maketitle

\begin{figure}
\includegraphics[width=4.5in]{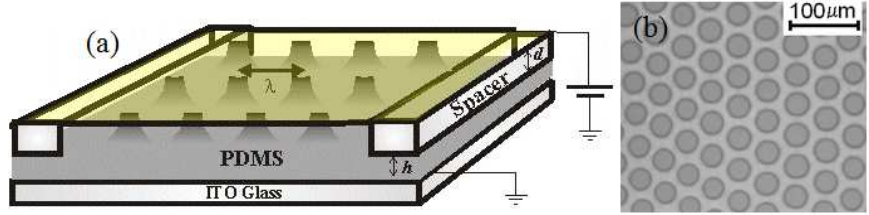}
\caption{ (a) Schematic diagram showing the experiment with PDMS film thickness, \textit{h} and air gap \textit{d} introduced by a spacer-pair. The Au coated top cover-glass contactor was typically placed with the Au-side facing up. (b) Hexagonally ordered circular columns formed in a viscoelastic liquid PDMS film with 1.5\% CL concentration, \textit{h} = 6.2 $\mu$m and V = 35 V.}
\label{fig:sch}
\end{figure}

The formation of self-organized instability structures in a thin liquid film \cite{ref1,ref2,ref3} and a soft solid polymer film \cite{ref4} is of great interest in diverse scientific and technological applications involving functional interfaces (adhesion, wetting, optics, etc.) and meso-patterning \cite{ref2,ref3}, especially with an applied electric field \cite{ref3}. Despite the fact that a majority of polymeric thin liquid and solid films are visco-elastic, the role of visco-elasticity or rheology in the spinodal pattern formation has not been systematically studied and remains unclear. Theoretically, it is understood that the instability length scale in a purely viscous liquid film has a long-wave character governed by a competition between the destabilizing field strength and surface tension \cite{ref1,ref2,ref3}. However, in a purely elastic solid, theory predicts a short-wave character with its length scale depending linearly on the film thickness, regardless of the field strength or elastic modulus \cite{ref4}. Isolating the effect of visco-elasticity in ultra-thin ($<$ 100 nm) liquid films destabilized by weak intermolecular interactions such as van der Waals is complicated by a host of other ill-defined factors \cite{ref5} such as glass transition, residual stresses, unknown details of intermolecular interactions, preparation conditions and nucleative (rather than spinodal) dewetting. On the other hand, previous studies on soft solid films are limited to contact instabilities induced by adhesion-debonding or peeling \cite{ref4,ref6,ref7} where lateral heterogeneities of the field and elasto-viscous fingering play significant roles. The objective of this work is to directly investigate the role of visco-elasticity or rheology in the process of spinodal pattern formation in a thin film destabilized by a field and to understand the transition from viscous to elastic behaviour. Towards this end, we employ an electric field \cite{ref3} to destabilize a polydimethyl siloxane film (Fig.~\ref{fig:sch}a) which is cross-linked to various degrees in order to uncover the entire spectrum of visco-elastic behaviour \cite{ref6} from a viscous liquid film displaying the long-wave instability to a soft elastomeric solid with a short wave instability and the nature of transition between the two regimes. The use of an electric field, rather than the van der Waals interaction, allows a precise tuning of the destabilizing force, which has to be made strong to induce instability of elastic films. Further, although the electrostatic field lithography \cite{ref3} has been shown to be a promising meso-patterning technique that has been widely applied to viscous melts, response of visco-elastic liquid and solids to an electric field has not been probed. Interestingly, a recent theoretical study points to the possibility of direct patterning of soft elastic solid film by an appropriately tuned electric field \cite{ref8}. Further, some theories and simulations of visco-elastic films \cite{ref9,ref10} suggest that the instability length scale, although different for liquid and solid films, are essentially independent of the precise rheology. In what follows, we report on the conditions for the pattern formation, length scale of patterns and their morphology while going seamlessly from films behaving as visco-elastic liquids to films behaving as soft visco-elastic solids. 

A wide variety of viscous and elastic films were prepared by varying the cross linker concentration in a commercially available two-part polydimethyl siloxane (PDMS) based elastomer, Sylgard 184 (Corning, USA). The cross linker (CL) percentage (which is usually about 10\% for the soft lithography stamps) was increased to obtain increasing ratio of $\mu$ (elastic storage modulus) to G$^{\prime\prime}$ (viscous loss modulus), both of which were characterized by a Bohlin Rheometer (Fig. 1 Supporting Material). The rheological response was essentially similar to that reported previously \cite{ref6} where the films with cross-linker concentration, CL $<$ 1\% ($\mu$/G$^{\prime\prime}$ $<$ 1) corresponded to a liquid regime, whereas the films acquired a substantial permanent, zero-frequency elastic modulus when CL was in excess of 2\% ($\mu$/G$^{\prime\prime}$ $>$ 1). 

The experimental setup for the electric field (EF) induced pattern formation involved placing the 
polymer film on the substrate electrode in the capacitor geometry with the counter parallel plate 
at a distance controlled by spacers as shown in Fig.~\ref{fig:sch}a. Transparent, conducting, 
pre-cleaned ITO coated glass slides were used as substrates on which the soft cross-linked PDMS films 
were spin coated from its solution in n-Hexane and then cross-linked. Films of thickness in the range 
of 1.5 $\mu$m - 75 $\mu$m were obtained by varying the spin coater speed (500 to 4800 rpm) and the 
polymer concentration. Films with 0\% to 3.5\% cross-linker were then annealed at $60\,^{\circ}\mathrm{C}$ 
for 4 hours or $110\,^{\circ}\mathrm{C}$ for 24 hours to obtain uniform films with surface 
roughness $<$ 10 nm. The counter electrode was in the form of a flexible cover glass (18 mm $\times$ 
18 mm, thickness $\approx$ 160 $\mu$m) or a rigid plate resting on a spacer (controlling the 
electrode-film distances, 0.03 $\mu$m $<$ \textit{d} $<$ 100 $\mu$m) made of a photoresist. The 
flexible contactor had a semi-transparent gold coating of thickness $\approx$ 50 nm on its top 
surface. A high voltage supply with controllable ramp-rate ($<$ 0.3 s)was used to apply the voltage 
between the two electrodes. The patterns were observed under a microscope and the images were 
captured using a digital camera. The additional voltage drop across the gold coated glass cover 
slip was accounted for in the calculations of the applied electric field.

\begin{figure}
\includegraphics[width=4.5in]{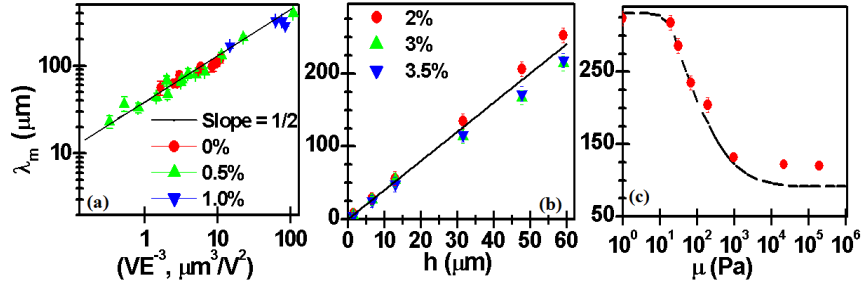}
\caption{ (a) Wavelength, $\lambda$ vs \textit{VE$^{-3}$} in log-log scale for viscoelastic liquid-like films for 0\% $<$ CL $<$ 1\%,  2.5 $\mu$m $<$ $h$ $<$ 75 $\mu$m, and $d$ = 30 nm, 2.5 $\mu$m, 4 $\mu$m, 20 $\mu$m , 40 $\mu$m, under bias conditions of  5 V $<$ $V$ $<$ 150 V. The line represents the viscous film theory ($\gamma$ = 19.8 mN/m and $\epsilon_p$ = 2.65). (b) $\lambda$ formed in solid-like viscoelastic films for different CL concentrations, the line represents the best fit with slope = 4.1. (c) Transition between the liquid-like and the solid-like regimes with predictions of eqn.~\ref{eq:one} for $h$ = 31.5 $\mu$m, $d$ = 50 $\mu$m; $V$ = 30 V for liquid-like films and $V_c$ for solid-like films.}
\label{fig:visco}
\end{figure}

For all visco-elastic films with a cross-linker concentration of 1\% or less, an array of circular 
columns packed locally in a hexagonal order appeared on the surface of film regardless of the voltage 
applied (Fig.~\ref{fig:sch}b). Their spacing remained constant from the earliest time an order could 
be detected on the film surface ($\sim$ 100 ms; Fig. 2 Supporting Material). Interestingly, the 
wavelengths of all these visco-elastic films could be described by a single master-curve for 
a \textit{purely} viscous liquid film \cite{ref3}, notwithstanding their elastic storage moduli 
(at CL = 1\%; low frequency modulus $\sim$ 20 Pa, 10 Hz modulus $\sim$ 1000 Pa). Fig.~\ref{fig:visco}a 
shows this master curve for a purely viscous liquid: $\lambda = C[VE^{-3}]^{1/2}$where, $C = 2\pi[2\gamma/\epsilon_p\epsilon_0(\epsilon_p-1)^{2}]^{1/2}$ and $E = V/[(d+h)\epsilon_p-h(\epsilon_p-1)]$. 
This observation is partly explained by a recent theory and simulations for visco-elastic liquids without 
a zero frequency elastic modulus \cite{ref9}, where $\lambda$ is determined solely by a competition 
between the destabilizing force and the surface tension, but not by the factors that influence the 
kinetics, such as high frequency elasticity and viscosity, both of which vary greatly in the CL 
range of 0-1 \%  (Fig. 1 Supporting Material). However, the dynamics of full pillar formation was 
seen to depend on the CL concentration or visco-elasticity.

However, if the CL concentration was increased beyond 2\%, an entire different scaling emerged as shown in Fig.~\ref{fig:visco}b. The onset of instability now required a minimum critical voltage (discussed later). The resulting pattern in these soft solid-like films was in fact indistinguishable from the case of liquid-like films with identical morphology of circular pillars packed in a hexagonal lattice. Interestingly, the wavelength in this regime becomes independent of the applied critical voltage, as well as the CL concentration or precise level of solid visco-elasticity. For these solid-like films, the wavelength exhibits a linear relationship with the film thickness with a best fit of $\lambda$ $\sim$ 4h (Fig.~\ref{fig:visco}b). This is reminiscent of the elastic contact instability seen in debonding and peeling of elastic adhesives \cite{ref4}. 

As shown in Fig.~\ref{fig:visco}c, the cross-over from the viscous liquid-like behavior to elastic solid-like behavior is continuous, although confined to a narrow range of elastic modulus. This is unlike the discontinuous first order transition found in a study of debonding of visco-elastic layers \cite{ref6}. The above observations can be understood from a more general linear stability analysis which gives the following dispersion relation for a visco-elastic thin film with constant viscous and elastic moduli \cite{ref10}:

\begin{equation}
(\eta\omega/\mu)= [-(\gamma/h\mu)q^2+(h/\mu)S](2qY(q))^{-1}-1
\label{eq:one}
\end{equation}

where, $q=kh$, $Y(q)=((1+e^{2q})^2+4e^{2q}q^2)(-1+e^{4q}-4e^{2q}q)^{-1}$ and $S=(\partial\pi/\partial h)=\epsilon_0\epsilon_pV^2(1-\epsilon_p)^2(\epsilon_pd+h)^{-3}$ where $\omega$ is the growth coefficient of instability, $k$ is its wave number, $\gamma$ is surface tension, $\mu$ is elastic shear modulus, $\eta$ is viscosity, $d$ is air gap, $h$ is film thickness, $\pi$ is excess electric pressure at the interface,  $\epsilon_0$ is dielectric permittivity of the free space, $\epsilon_p$ is dielectric constant of the polymer (= 2.65) and $V$ is applied voltage. From the above relation, the liquid-like scaling is obtained when the parameter ($\gamma$/$h$$\mu$) is large, whereas the solid-like scaling $\lambda$ $\sim$ 3$h$ is recovered for small ($\gamma$/$h$$\mu$).

Fig.~\ref{fig:visco}c shows the wavelength obtained from eqn.~\ref{eq:one} which adequately describes the features of transition from the liquid-like behavior at low ($<$$\sim$ 10 Pa) elastic modulus to elastic solid-like behavior at moderately high values ($>$$\sim$ 1000 Pa). Interestingly, the wavelength in all cases is independent of viscosity as predicted by eqn.~\ref{eq:one}. Further, in both the liquid-like and the solid-like regimes, it also becomes independent of the elastic modulus, which remains important only in the transition regime. Basically, in the liquid-like regime, surface tension is the dominant stabilizing mechanism, whereas in the solid-like regime, the elastic strain dominates. The transition is thus governed both by surface tension and elasticity in the form of the parameter, ($\gamma$/$h$$\mu$). Further, in the solid-like regime, $\lambda$ is solely governed by the pattern that minimizes the elastic energy penalty \cite{ref4,ref8,ref10} and the film viscosity merely governs the dynamics of pattern formation \cite{ref10}.  The solid-regime scaling may be physically understood by nothing that a simple scaling relation for the elastic strain energy (per unit area) of the film is given by \cite{ref4}: \textit{U} $\sim$ $\mu$\textit{h}$\delta$$^{2}$($\lambda$$^{-1}$+$\lambda$\textit{h}$^{-2}$)$^{2}$, where $\mu$ is the elastic shear modulus, \textit{h} is film thickness and $\delta$ is the vertical amplitude of the pattern. Basically, the energy penalty is high both for very short ($\lambda$$<<$\textit{h}) and for very long ($\lambda$$>>$\textit{h}) waves, and thus the minimum elastic energy pattern (for which $\partial$\textit{U}/$\partial\lambda$=0) has a length scale of the order of the film thickness, $\lambda$ $\sim$ \textit{h}. Thus, unlike the case of a liquid-like film, instability in a visco-elastic solid-like film is independent of the nature of the destabilizing force, as long as it exceeds a critical value to overcome the elastic stiffness of the film. For a rigid top electrode, the critical voltage, \textit{V$_{c}$} for the onset of instability may be obtained from eqn.~\ref{eq:one} and is given by \cite{ref7,ref8}:

\begin{equation}
 V_c^2=
 \frac{6.22\mu}{\epsilon_p\epsilon_0(\epsilon_p-1)^{2}}
  \frac{(\epsilon_pd+h)^{3}}{h}
\label{eq:two}
\end{equation}

\begin{figure}
\includegraphics[width=4.5in]{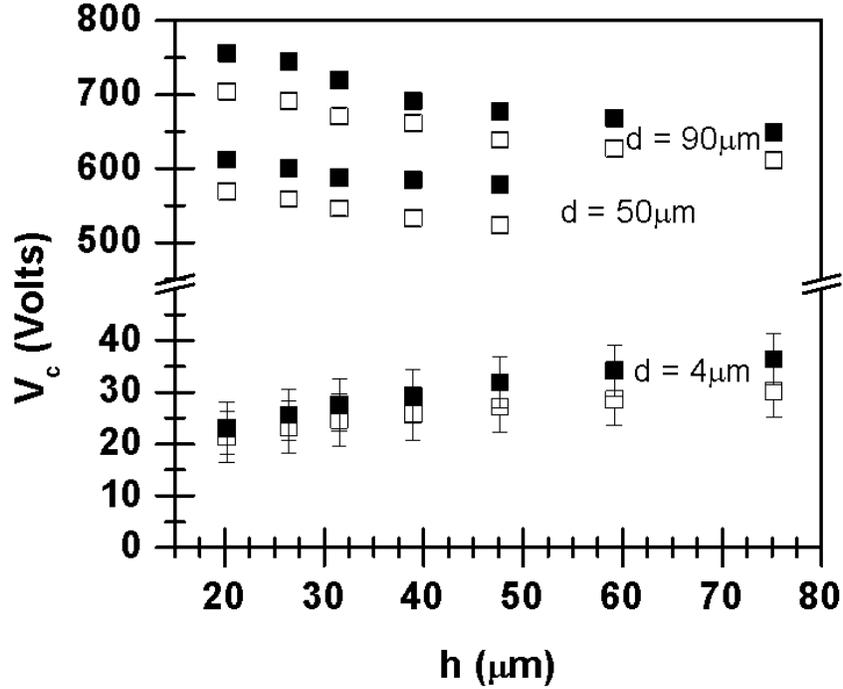}
\caption{ Dependence of critical voltage, \textit{V$_{c}$} on film thickness, \textit{h} for different values of air gap, \textit{d}. The hollow and solid symbols represent CL concentrations of 3\% and 3.5\% respectively.}
\label{fig:volt}
\end{figure}

It was indeed observed that the onset of instability in solid films required a critical voltage and pillars were not formed when \textit{V} was maintained marginally lower than the \textit{V$_{c}$} even for several hours. The pillars formed at \textit{V $>$ V$_{c}$} could extend and contact the flexible-thin contactor for the entire range of experiments where \textit{d} varied from 2.5 $\mu$m to 90 $\mu$m. The trend of the \textit{V$_{c}$} values predicted by eqn.~\ref{eq:two} with respect to \textit{h} is observed qualitatively in our measurements shown in Fig.~\ref{fig:volt}. When \textit{d} is small compared to \textit{h}, \textit{V$_{c}$} increases with \textit{h}. However, for large values of \textit{d}, the \textit{V$_{c}$} scales inversely with \textit{h}. Notwithstanding these qualitative trends, eqn.~\ref{eq:two}, with $\mu$ = 0.1 MPa, quantitatively predicts voltages that are several times larger than the experimental values and are in fact large enough to cause a dielectric breakdown of the polymer and air! In our experiments, we could observe elastic pillar formation only when either a flexible top electrode was used where a small bending ($\sim$ 2 $\mu$m at 500 V; much smaller than the air gap) of the electrode at its center was observed or slightly non-parallel rigid electrode was used. This small bending of the electrode and the lateral field gradient thus created is an essential factor in kick-starting the elastic instability which is a nucleation phenomena at much lower voltages. However, the length scale of instability ($\approx$ 4\textit{h}) and the pattern morphology (hexagonally packed circular pillars) were still as predicted in simulations involving rigid electrodes \cite{ref8}. 

\begin{figure}
\includegraphics[width=4.5in]{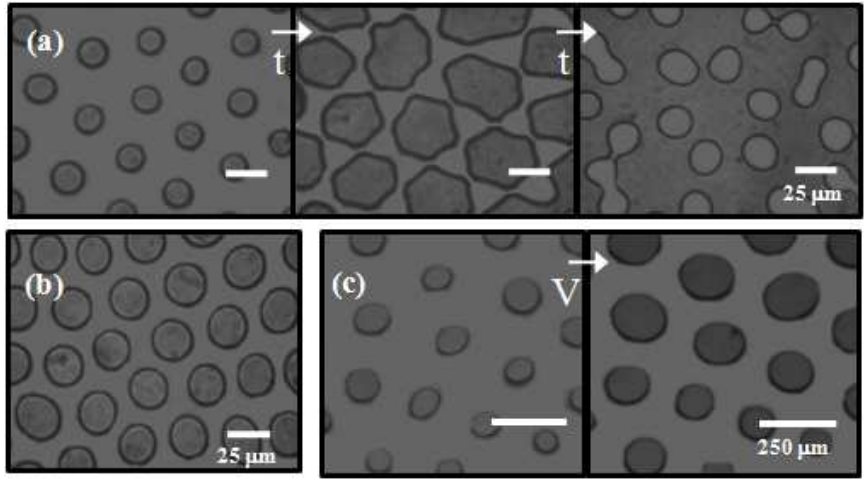}
\caption{ (a) The growth of the hexagonally ordered pillars into isolated voids for 1.5\% CL concentration, \textit{h} = 3 $\mu$m, \textit{d} = 2.5 $\mu$m and \textit{\textit{d}} = 60 Volts at time, \textit{t} $\approx$ 0 s, \textit{t} $\approx$ 5 s, \textit{t} $\approx$ 65 s. (b) Optical Micrograph of elastic PDMS film surface with 3\% CL concentration and \textit{h} = 6.6 $\mu$m, \textit{d} = 2.5 $\mu$m for \textit{\textit{V}} = 97 V($\approx$\textit{V$_{c}$}). (c) Pillar size growth for 3.5\% CL concentration (\textit{h} = 75.2 $\mu$m and \textit{d} = 50 $\mu$m) for external voltage \textit{V} = 630 V ($\approx$ \textit{$V_{c}$}) and when \textit{V} = 850 V (right side) with a slew rate $<$ 0.3 s.}
\label{fig:pics}
\end{figure}

Finally, we briefly present some other notable distinctions between the behavior of visco-elastic liquid-like and solid-like films. In the transition regime, mixed signatures of both the regimes were observed. For example, 1.5\% CL films required a critical voltage for the onset of instability as in the solid-like regime, but showed liquid-like ripening of structures at longer times. Scaling of wavelength with the film thickness, $\lambda$ = $nh$, with $n$ $>$ 4, is also found intermediate to the liquid-like and the solid-like films and could be predicted by eqn.~\ref{eq:one} (Fig. 3 Supporting Information). In liquid-like films, pillars formed initially coalesce rapidly, thus altering the pattern dimensions, density and geometry as shown in Fig.~\ref{fig:pics}a for a transition regime visco-elastic film at 1.5\% CL. Eventually, the pillar structure (polymer-in-air) is transformed into a pattern of voids surrounded by the polymer (last picture in Fig.~\ref{fig:pics}a). This morphological phase inversion was also predicted in simulations of purely viscous films \cite{ref11}. The kinetics of pillar formation and their coalescence could be hastened by increasing voltage and decreasing viscosity. Upon removal of the electric field, pillar patterns disappeared, but the inverted void-patterns remained robust because of their greater area of adhesion. The late time pattern ripening and morphological inversion extended well into solid-like regime, albeit at much slower rates, when the elastic and loss moduli were comparable ($\mu/G^{\prime\prime}$ $<$ 5; CL $<$ 2.5\%; frequency range - 0.1 Hz to 10 Hz). In that part of the transition regime that shows a solid-like behaviour for the wavelength (1.5\% $<$ CL $<$ 2.5\%), the film behaves elastically at short times but flows in long times. In contrast, in dominantly elastic films (CL $>$ 2.5\%), coalescence of the pillars was not observed even after long times (Fig.~\ref{fig:pics}b). An abrupt increase in \textit{V} ($>$ \textit{V$_{c}$}) for these elastic films merely increased the pillar contact radius while maintaining a constant $\lambda$ (Fig.~\ref{fig:pics}c). Upon removal of EF, a large hysteresis in the form of persistence of the pillar pattern was observed in the elastic regime.

In summary, a visco-elastic film subjected to an external destabilizing field displays two clearly distinct regimes of instability which have wavelengths corresponding to a purely viscous liquid and a purely elastic solid, regardless of their detailed rheology. The transition between the two regimes of long and short waves is confined to a narrow zone of a parameter, ($\gamma$/$h$$\mu$). Wavelength in the liquid-like regime depends on the applied field, film thickness and surface tension in a nonlinear fashion, whereas the wavelength in the solid-like elastic regime depends linearly on the film thickness independent of the field strength and material properties. However, the kinetics of instability is influenced by rheology. An important difference between a liquid-like film and a solid-like film is the presence of a critical voltage required to engender the pillar formation in the latter case. These observations are of general interest in the instability, pattern formation and self-organized meso-patterning in thin visco-elastic films under the influence of destabilizing forces such as electric fields and intermolecular interactions.


\begin{thebibliography}{99}
\bibitem{ref1} 
			G. Reiter, Phys. Rev. Lett. {\bf 68}, 75 (1992).
      A. Oron, S. H. Davis and S. G. Bankof, Rev. Mod. Phys. {\bf 69}, 931 (1997) 
      R. Xie, A. Karim, J. F. Douglas, C. C. Han and R. A. Weiss, Phys. Rev. Lett. {\bf 81}, 1251 (1998).
      A. Sharma and R. Khanna, Phys. Rev. Lett. {\bf 81}, 3463 (1998).
			U. Thiele, M. G. Velarde, K. Neuffer, Phys. Rev. Lett. {\bf 87}, 016104 (2001).
			J. Becker, {\it et al.} Nat. Mater. {\bf 2}, 59 (2003).

\bibitem{ref2} 
			S. Y. Chou, L. Zhuang and L. Guo, Appl. Phys. Lett. {\bf 75}, 1004 (1999).
  		R. Konnur, K. Kargupta and A. Sharma, Phys. Rev. Lett. {\bf 84}, 931 (2000).
			Z. Zhang, Z. Wang, R. Zing and Y. Han, Polymer {\bf 44}, 3737 (2003).
			S. Seo, J. Park and H. H. Lee, Appl. Phys. Lett. {\bf 86}, 133114 (2005).     
			A. Sehgal, V. Ferreiro, J. F. Douglas, E. J. Amis and A. Karim, Langmuir {\bf 18}, 7041 (2002).

\bibitem{ref3}
			E. Schaffer, T. Thurn-Albrechet, T. P. Russell and U. Steiner, Nature {\bf 403}, 874 (2000).
			E. Schaffer, T. Thurn-Albrechet, T. P. Russell and U. Steiner, Europhys. Lett. {\bf 53}, 518 (2001).
			M. D. Morariu, {\it et al.} Nat. Mater. {\bf 2}, 48 (2003).
			S. Harkema and U. Steiner, Adv. Funct. Mater. {\bf 15}, 2016 (2005).
			N. Wu, L. F. Pease III and W. B. Russel, Adv. Funct. Mater. {\bf 16}, 1992 (2006).
			N. E. Voicu, S. Harkema and U. Steiner, Adv. Funct. Mater. {\bf 16}, 926 (2006).
			M. D. Dickey, {\it et al.} Chem. Mater. {\bf 18}, 2043 (2006).
  			N. E. Voicu, S. Ludwigs and U. Steiner, Adv. Mater. {\bf 20}, 3022 (2008).

\bibitem{ref4}
			A. Ghatak, M. K. Chaudhury, V. Shenoy and A. Sharma, Phys. Rev. Lett, {\bf 85}, 4329 (2000).
			V. Shenoy and A. Sharma, Phys. Rev. Lett. {\bf 86}, 119 (2001).
			W. Monch and S. Herminghaus, Europhys. Lett. {\bf 53}, 525 (2001).
			A. Ghatak and M. K. Chaudhury, Langmuir {\bf 19}, 2621 (2003).
			J. Sarkar, V. Shenoy and A. Sharma, Phys. Rev. Lett. {\bf 93}, 018302 (2004).
			M. Gonuguntla, {\it et al.}, Phys. Rev. Lett. {\bf 97}, 018303 (2006).
			M. Gonuguntla, A. Sharma, R. Mukherjee, S. A. Subramanian, Langmuir {\bf 22}, 7066 (2006).

\bibitem{ref5}
			J. L Keddie and R. A. L. Jones, R. A. Cory, Europhys. Lett. {\bf 27}, 59 (1994).
			G. Reiter and A. Sharma, Euro. Phys. J. E {\bf 12}, 397 (2003).
			M. Hamieh, {\it et al.}, Nat. Mater. {\bf 4}, 754 (2005).

\bibitem{ref6}
			J. Nase, A. Lindner and C. Creton, Phys. Rev. Lett. {\bf 101}, 074503 (2008).

\bibitem{ref7}
			N. Arun, A. Sharma, V. Shenoy and K. S. Narayan, Adv. Mater. {\bf 18}, 660 (2006).

\bibitem{ref8}
			J. Sarkar, A. Sharma and V. Shenoy, Phys. Rev. E {\bf 77}, 031604 (2008).

\bibitem{ref9}
			L. F. Pease and W. B. Russel, J. Non-Newtonian Fluid Mech. {\bf 102}, 233 (2002).
			R. V. Craster and O. K. Matar, Phys. Fluids {\bf 17}, 032104 (2005).
			L. Wu and S. Y. Chou, J. Non-Newtonian Fluid Mech. {\bf 125}, 91 (2005).
			G. Tomar, V. Shankar, A. Sharma and G. Biswas, J. Non-Newtonian Fluid Mech. {\bf 143}, 120 (2007).
			G. Tomar, V. Shankar, A. Shukla, A. Sharma and G. Biswas, Eur. Phys. J. E-Soft Matter {\bf 20}, 185 (2006).

\bibitem{ref10}
			V. Shenoy and A. Sharma, J. Mech. and Phys. Solids {\bf 50}, 1155 (2002).

\bibitem{ref11}
			R. Verma, A. Sharma, K. Kargupta and J. Bhaumik, Langmuir {\bf 21}, 3710 (2005).
\end{thebibliography}
\end{document}